\documentclass[twoside]{LCWS11}
\usepackage[latin1]{inputenc}
\usepackage[dvips]{graphicx,epsfig,color}
\usepackage{wrapfig,rotating}
\usepackage{amssymb,amsmath,array}
\usepackage{cite}

\input paperdef

\newcommand{\DecayCmNW}[2]{\cham{#1} \to \neu{#2} W^-}

\newcommand{\SN}{${\cal S}$}
\newcommand{\SE}{\ensuremath{{\cal S}_>}}
\newcommand{\SZ}{\ensuremath{{\cal S}_<}}

\begin{document}

\title{
Neutralinos from Chargino Decays \\ in the Complex MSSM
} %% 
%***********************************************************************
% AUTHORS INFORMATION AREA
%***********************************************************************
\author{S.~Heinemeyer$^{1}$%
, F.~v.~d.~Pahlen$^{1}$%
~and C.~Schappacher$^{2}$
\thanks{KA-TP-05-2012}
\vspace{.3cm}\\
$1$-Instituto de F\'isica de Cantabria (CSIC-UC)%, 
\\
E--39005 Santander, Spain
\vspace{.1cm}\\
$2$-Institut f\"ur Theoretische Physik, Karlsruhe Institute of Technology, 
\\
D--76128 Karlsruhe, Germany
}
%%***********************************************************************
% END OF AUTHORS INFORMATION AREA
%***********************************************************************

\maketitle

\begin{abstract}
   We review the evaluation of two-body decay modes of charginos in the 
   Minimal Supersymmetric Standard Model with complex parameters (cMSSM). 
   Assuming heavy scalar quarks we take into account all decay channels 
   involving charginos, neutralinos, (scalar) leptons, Higgs bosons and SM gauge bosons. 
   The evaluation of the decay widths is based on a full one-loop calculation
   including hard and soft QED radiation. 
   Here we focus on the decays involving the Lightest Supersymmetric Particle (LSP), 
   i.e.\ the lightest neutralino, 
   or a heavier neutralino and a $W$~boson. 
   The higher-order corrections of the chargino decay widths can easily reach a level of $\pm 10\%$,
   translating into corrections of similar size in the respective branching ratios.
   These corrections are important for the correct interpretation of
   LSP and heavier neutralino production at the LHC and at a future linear $e^+ e^-$ collider.
\end{abstract}

\section{Introduction}
{The search for physics effects beyond the Standard Model (SM), 
both at present and future colliders,
constitutes one of the priorities of current high energy physics,
where the 
Minimal Supersymmetric Standard Model (MSSM)~\cite{mssm} is one of the leading candidates.
 }
A  related important task is investigating the production and measurement of 
the properties of Cold Dark Matter (CDM).
The MSSM offers a natural candidate of CDM,
the Lightest Supersymmetric Particle (LSP), 
i.e.\ the lightest neutralino,~$\neu{1}$~\cite{EHNOS}.
Having a stable LSP also ensures that any produced supersymmetric particle 
will lead to cascades with neutralinos in the final state, 
motivating experimental and phenomenological analyses of these decay chains.
While discoveries of supresymmetric particles will possibly be made by the LHC, 
a precise determination of their properties  is expected at the ILC~\cite{teslatdr,ilc,lhcilc}
 (or any other future $e^+ e^-$ collider such as CLIC).

Charginos, $\cha{i}$, ($i=1,2$), and neutralinos, $\neu{j}$, ($j=1,2,3,4$), are, respectively,  
the charged and neutral supersymmetric partners of the Higgs and gauge bosons.
Therefore masses and couplings of charginos and neutralinos depend on common parameters,
and 
an analysis of chargino decays provides
direct and indirect information on the neutralino sector.

In order to yield a sufficient accuracy, one-loop corrections to
the various chargino decay modes have to be considered.
A precise calculation of the branching ratio (BR) at the one-loop level 
requires the calculation of all decay modes at this level of precision.
Here we review the results for the evaluation of these decay modes (and BRs)
obtained in the MSSM with complex parameters (cMSSM)~\cite{ChaPaper}
(original results for the tree-level decays were presented 
in \cite{chadectree}).
We show results for
\begin{align}
&  \Gamma(\cha{2} \to \neu{j} W^\pm),\quad j = 1,2,3~.
\end{align}
The total decay width is defined as the sum of all the partial two-body decay widths, 
{\it all} evaluated at the one-loop level.
Detailed references to existing calculations of these decay widths, branching ratios, 
as well as about the extraction of complex phases can be found in \citere{ChaPaper}.
Our results will be implemented into the Fortran code {\tt FeynHiggs}~\cite{feynhiggs,mhiggslong,mhiggsAEC,mhcMSSMlong}.

% %%%%%%%%%%%%%%%%%%%%%%%%%%%%%%%%%%%%%%%%%%%%%%%%%%%%%%%%%%%%%%%%%%%%%%%%%%%%%

\section{Renormalization of the cMSSM}

All the relevant two-body decay channels have been evaluated at the one-loop level, including hard QED radiation.
This requires the simultaneous renormalization of several sectors of the cMSSM: the gauge and Higgs sector,
the chargino/neutralino sector, and the lepton and slepton sector.
The on-shell renormalization conditions for the chargino/neutralino sector are fixed
requiring that the masses of the two charginos and of the lightest neutralino are not renormalized.
An analysis of various renormalization schemes for the chargino/neutralino sector was recently
published in \citere{onshellCNmasses}.
Further details about our notation and about the renormalization of the cMSSM 
can be found in \citeres{Stop2Paper,ChaPaper,SbotRen}.

In order to highlight the important role of the absorptive contributions in the presence of complex couplings,
we also evaluated for comparison the decay widths neglecting the imaginary parts 
of self-energy type corrections to external (on-shell) particles.
These imaginary contributions, in product with an imaginary part of a complex coupling (such as $\MOne$ in our case),
can give an additional real contribution to the decay width. 
This contribution 
is odd under charge conjugation and
leads to a difference in the
decay widths for the chargino and its antiparticle.
The resulting CP-asymmetry, however, is one-loop suppressed (and will not be analyzed here).

The diagrams and corresponding amplitudes have been obtained with 
\fa~\cite{feynarts}. 
The model file, including the MSSM counter terms, 
is based largely on \citere{dissTF} and is discussed in more detail in \citere{Stop2Paper}. 
The further evaluation has been performed with 
\fc\ (and \looptools)~\cite{formcalc}. 
As regularization scheme for the UV-divergences we
have used constrained differential renormalization~\cite{cdr}, 
which has been shown to be equivalent to 
dimensional reduction~\cite{dred} at the \onel\ level~\cite{formcalc}. 
Thus the employed regularization preserves SUSY~\cite{dredDS,dredDS2}. 
All UV-divergences cancel in the final result.
(Also the IR-divergences cancel
in the one-loop result as required.)

% %%%%%%%%%%%%%%%%%%%%%%%%%%%%%%%%%%%%%%%%%%%%%%%%%%%%%%%%%%%%%%%%%%%%%%%%%%%%%

\section{Numerical results}
\label{sec:figures}
The numerical examples shown below have been evaluated using the parameters given in \refta{tab:para}.
We assume that the scalar quarks are heavy such that they do not contribute to the total decay widths
of the charginos.
We invert the expressions of the chargino masses in order to express the parameters 
$\mu$ and $\MTwo$
(which are chosen real)
as a function of $\mcha{1}$ and  $\mcha{2}$.
This leaves two choices for the hierarchy of $\mu$ and $\MTwo$:
\begin{align}
\label{eq.SE}
\SE &: \mu > \MTwo \quad (\cha{2} \mbox{~more higgsino-like})~, \\
\label{eq.SZ}
\SZ &: \mu < \MTwo \quad (\cha{2} \mbox{~more gaugino-like})~.
\end{align}

%%%%%%%%%%%%%%%%%%%%% T A B L E %%%%%%%%%%%%%%%%%%%%%%%%%%%%%%%%%%%%%%%%%%%%%%
\begin{table}[ht!]
\renewcommand{\arraystretch}{1.5}
\BC
\begin{tabular}{|c||c|c|c|c|c|c|c|c|}
\hline
Scen.\ & $\tb$ & $\MHp$ & $\mcha{2}$ & $\mcha{1}$ 
       & $\MslL$ & $\MslR$ & $\Al$ 
\\ \hline\hline
\SN & 20 & 160 & 600 & 350 & 300 & 310 & 400 
\\ \hline
\end{tabular}
\caption{MSSM parameters for the initial numerical
  investigation; all 
  masses are in GeV. 
}
\label{tab:para}
\EC
\renewcommand{\arraystretch}{1.0}
\end{table}
%%%%%%%%%%%%%%%%%%%%% T A B L E %%%%%%%%%%%%%%%%%%%%%%%%%%%%%%%%%%%%%%%%%%%%%%

The absolute value of $\MOne$ is fixed via the GUT relation (with $|\MTwo|\equiv \MTwo $),
\begin{align}
|\MOne| &= \frac{5}{3} \tan^2 \thw \MTwo \approx \edz \MTwo~.
\label{M1M2}
\end{align}
leaving $\phiMe$ as a free parameter.

 The values of $\mcha{1,2}$ allow copious 
production of the charginos in SUSY cascades at the LHC.
Furthermore, the production of $\cha{1}\champ{2}$ or $\chap{1}\cham{1}$
at the ILC(1000), i.e.\ the ILC with $\sqrt{s} = 1000 \gev$, 
via 
$e^+e^- \to \cha{1}\champ{1,2}$ will be possible,
with all the subsequent decay modes to a neutralino and a $W$~boson 
being open. 
The clean environment of the ILC would
permit a detailed study of the chargino decays~\cite{ilc,lhcilc}.
For the values in \refta{tab:para} and unpolarized beams
we find, for $\SE$ ($\SZ$),
$\si(e^+e^- \to \cha{1}\champ{2}) \approx 4\, (12)~{\rm fb}$, and
$\si(e^+e^- \to \chap{1}\cham{1}) \approx 55\, (80)~{\rm fb}$. 
Choosing appropriate polarized beams these cross sections can be
enhanced by a factor of approximately $2$ to $3$.
An integrated luminosity of $\sim 1\, \iab$ would yield about 
$4-12 \times 10^3$ $\cha{1}\champ{2}$ events and about
$55 - 80 \times 10^3$ $\chap{1}\cham{1}$ events, with appropriate
enhancements in the case of polarized beams.

The ILC environment would result in an accuracy of
the relative branching ratio 
close to the statistical
uncertainty: 
assuming an integrated luminosity of  $1\, \iab$ 
a BR of $10$\% could be determined to $\sim 2\%$ for the $\mcha{i}$ values of \refta{tab:para}. 

\begin{figure}[htb!]
\begin{center}
\begin{tabular}{c}
\includegraphics[width=0.49\textwidth,height=4.6cm]{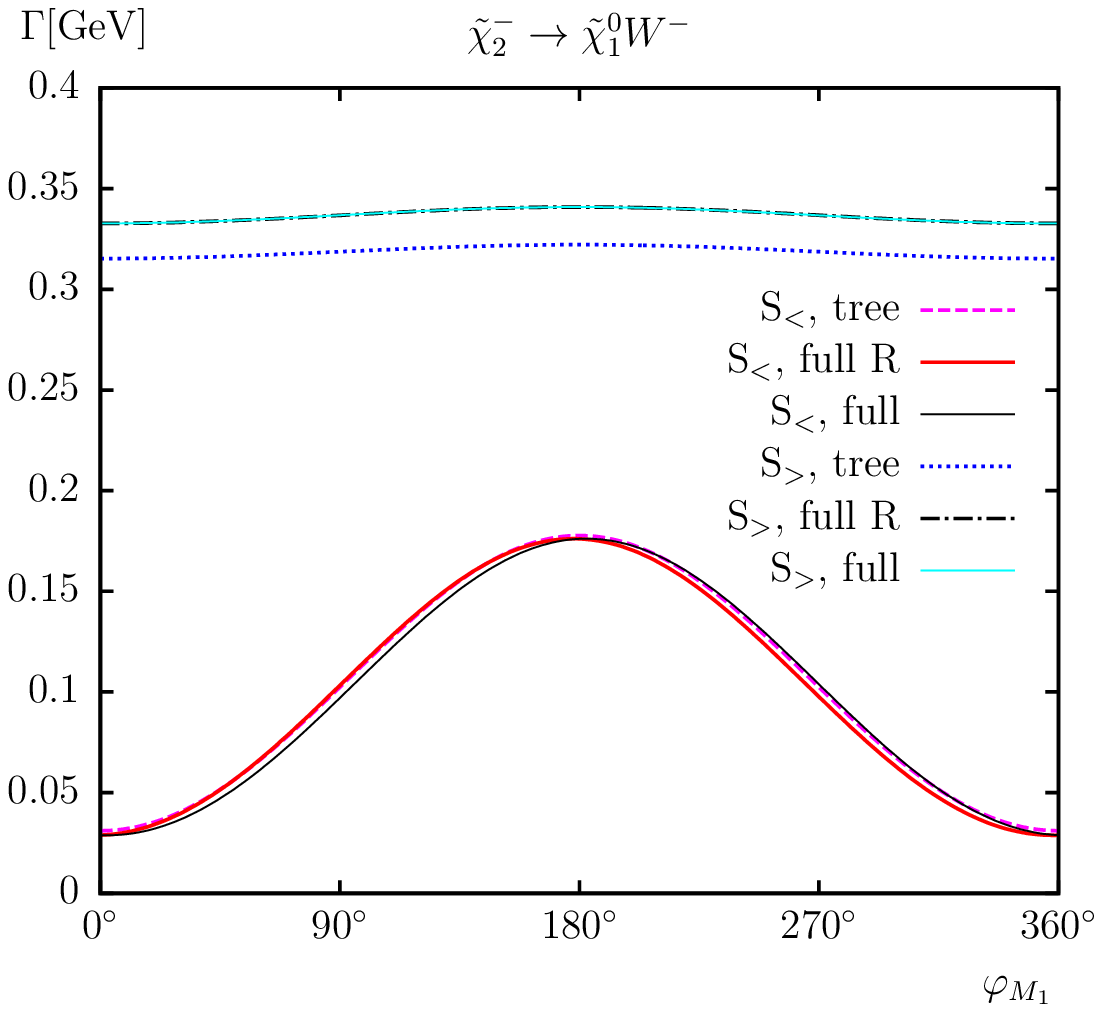}
\hspace{-4mm}
\includegraphics[width=0.49\textwidth,height=4.6cm]{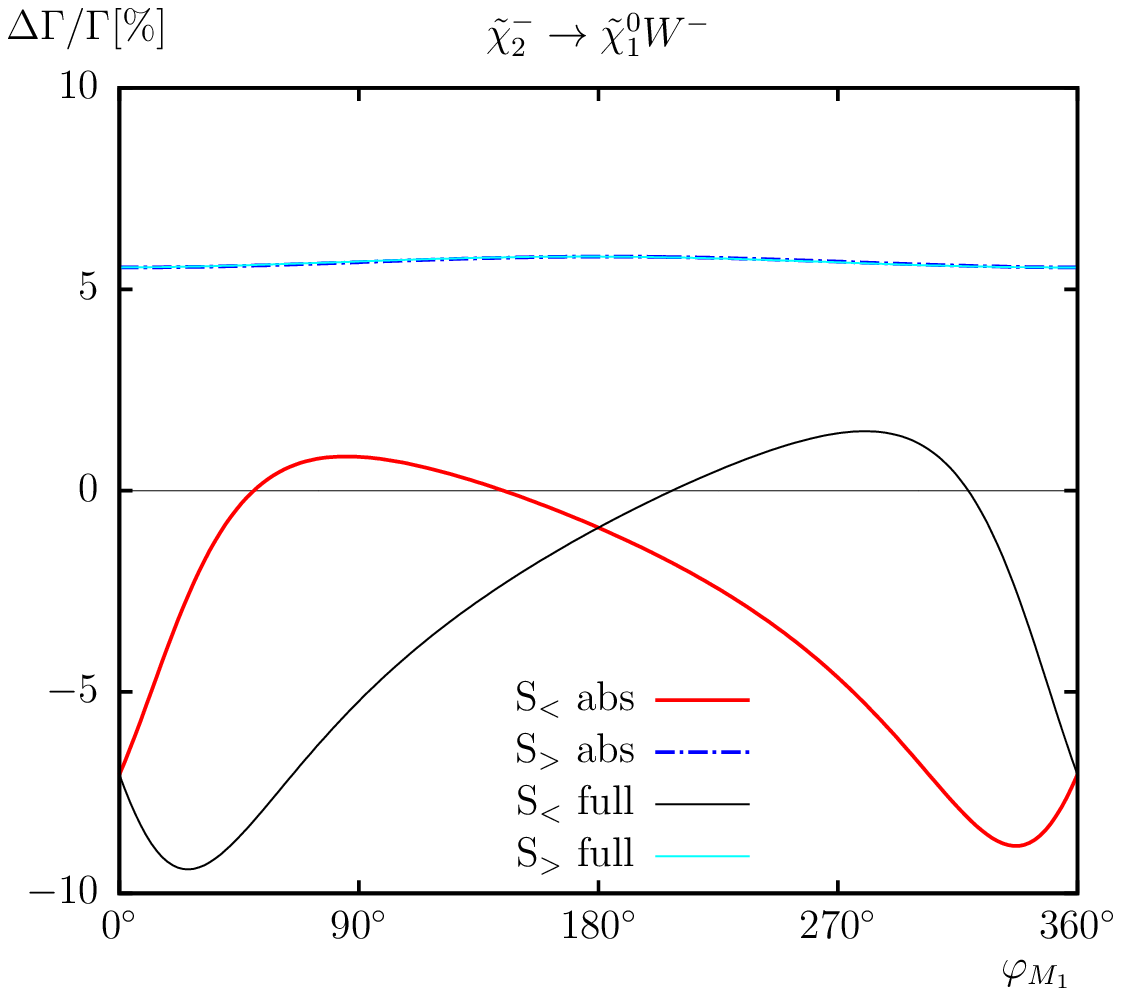} 
\end{tabular}
%\vspace{2em}
\begin{tabular}{c}
\includegraphics[width=0.49\textwidth,height=4.6cm]{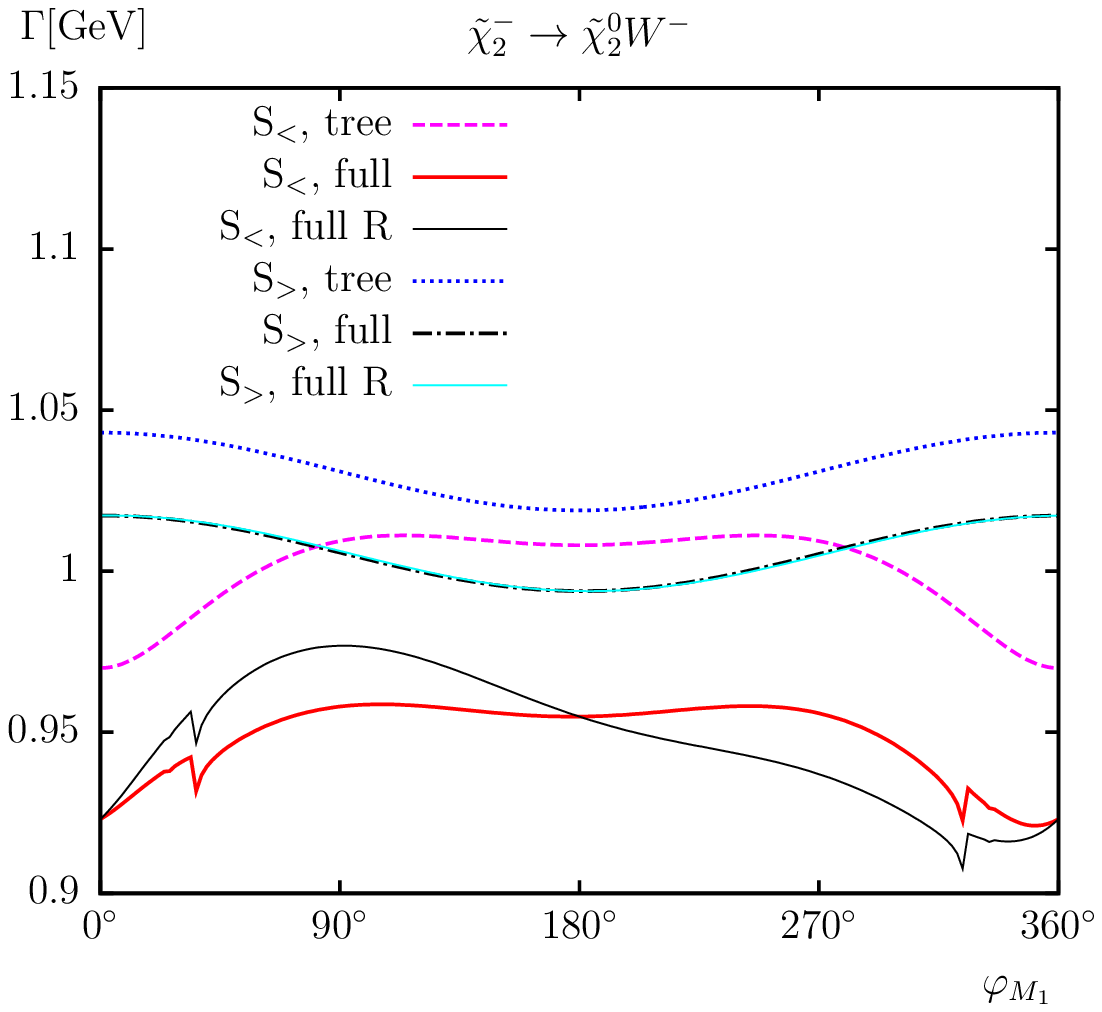}
\hspace{-4mm}
\includegraphics[width=0.49\textwidth,height=4.6cm]{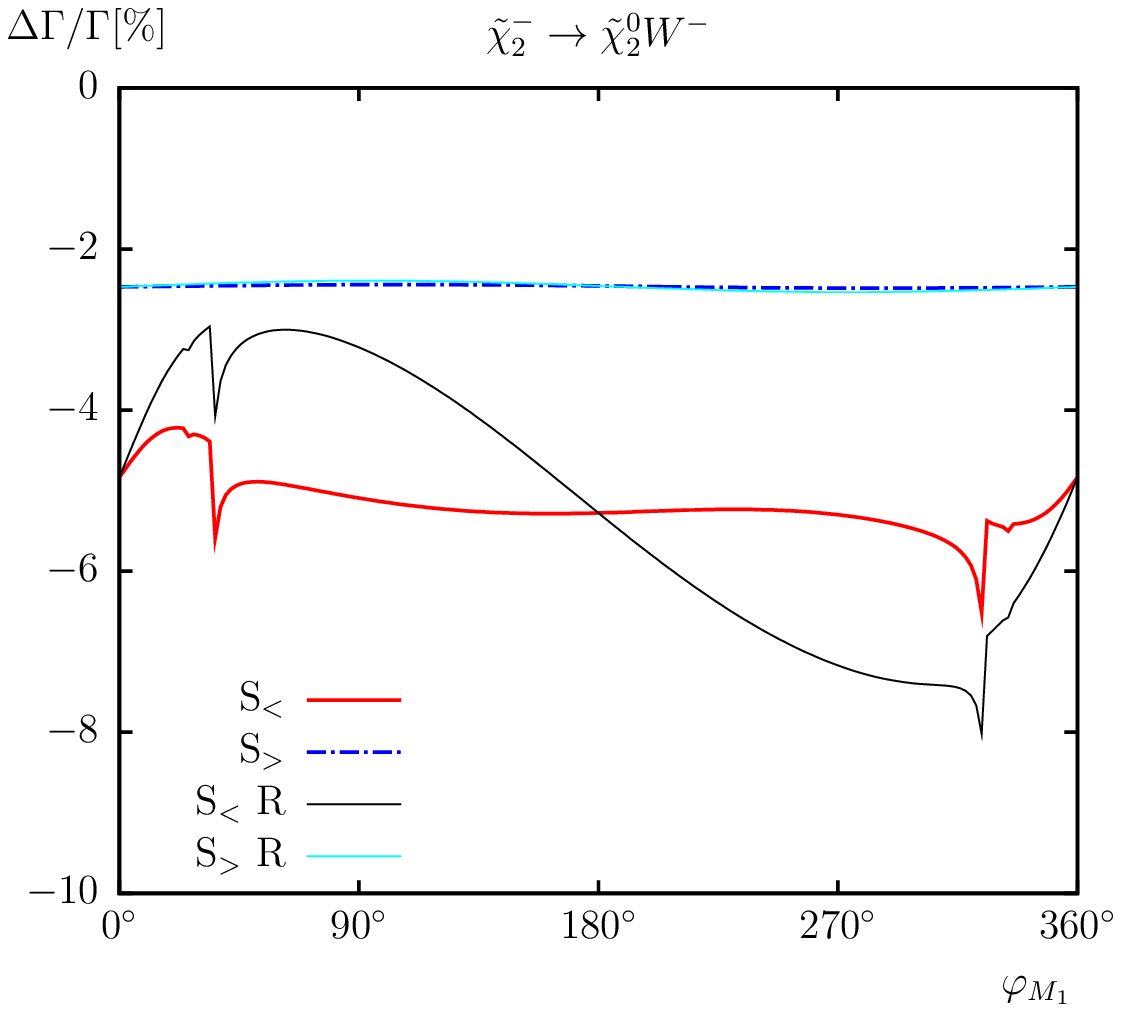} 
\end{tabular}
%\vspace{2em}
\begin{tabular}{c}
\includegraphics[width=0.49\textwidth,height=4.6cm]{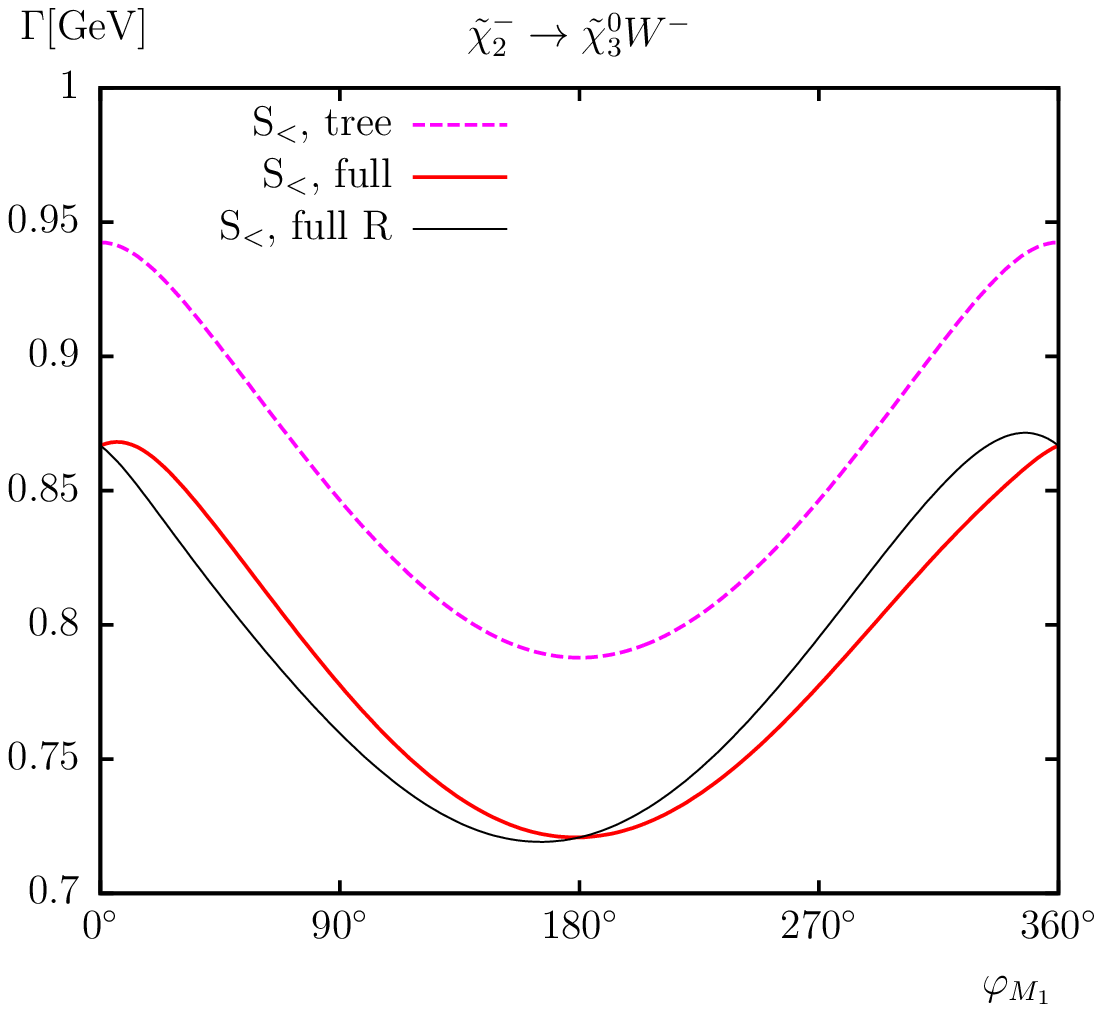}
\hspace{-4mm}
\includegraphics[width=0.49\textwidth,height=4.6cm]{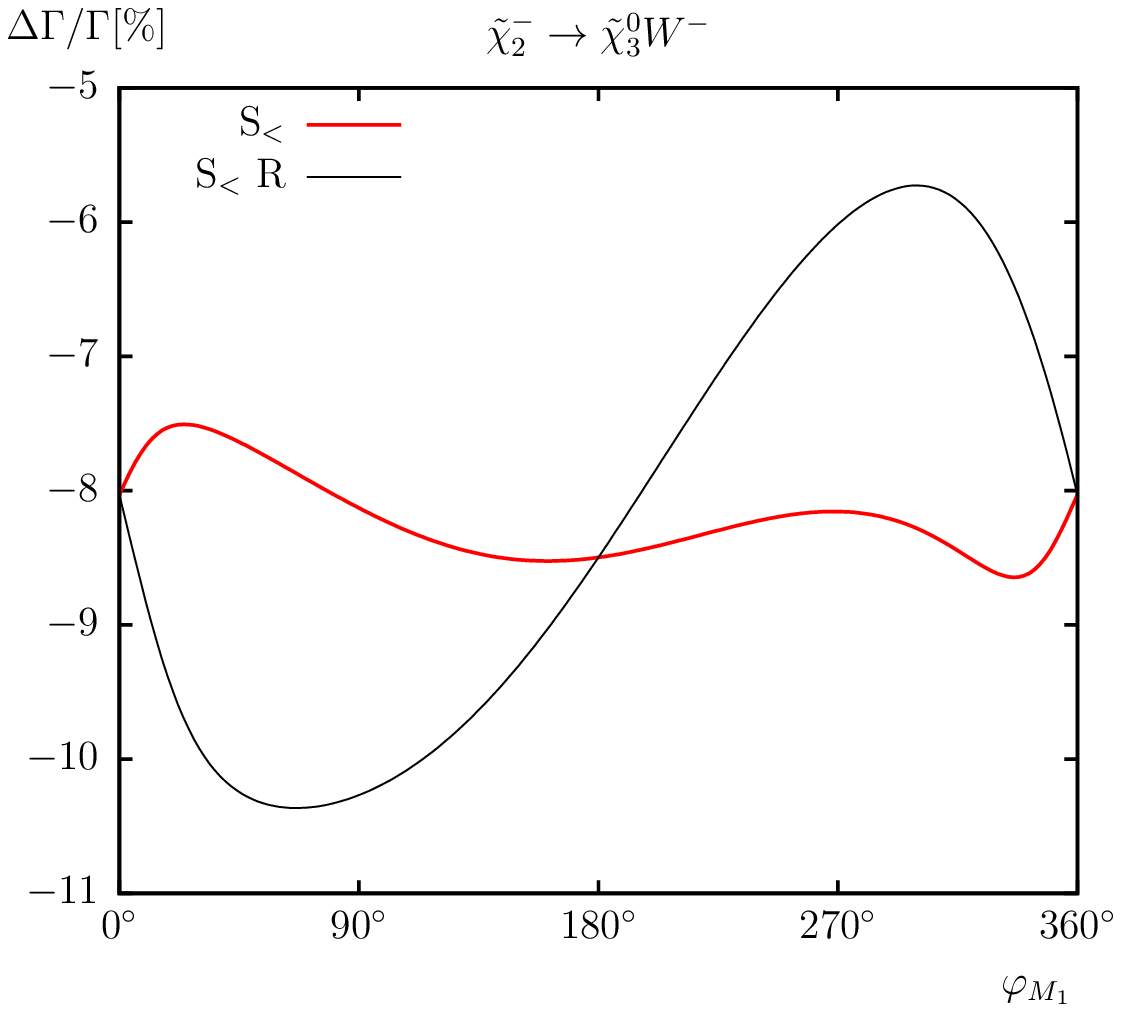} 
\end{tabular}
%\vspace{-2em}
\caption{
  $\Ga(\DecayCmNW{2}{j})$, $j=1,2,3$.
  Tree-level (``tree'') and full one-loop (``full'') corrected 
  decay widths are shown with the parameters chosen according to \SN\
  (see \refta{tab:para}), with $\phi_{\MOne}$ varied.
  Also shown are the full one-loop corrected decay widths omitting
  the absorptive contributions (``full R'').
  The left (right) plots show 
  (the relative size of the corrections of)
  the decay width.
}
\label{fig:PhiM1.cha2neujw}
\end{center}
\end{figure}

The results shown here consist of ``tree'', which denotes the tree-level 
value and of ``full'', which is the decay width including {\em all} one-loop 
corrections. Also shown in Fig.~\ref{fig:PhiM1.cha2neujw}, 
is the result leaving out the contributions from absorptive 
parts of the one-loop self-energy corrections as discussed in the previous section,
 labeled as ``full R''. 
Not shown here are the BRs and their relative corrections, since they are more parameter dependent.

In Figure \ref{fig:PhiM1.cha2neujw} we show 
  $\Ga(\DecayCmNW{2}{1})$ (top), 
  $\Ga(\DecayCmNW{2}{2})$ (middle), 
and  $\Ga(\DecayCmNW{2}{3})$ (bottom row)
as a function of of $\phiMe$, for the parameters of  \refta{tab:para}.
The left (right) columns display the  (relative one-loop correction to the) decay width.

We observe a strong dependence on $\phiMe$ in scenario $\SZ$, 
in which the three lightest neutralinos are highly mixed states.
The effect of the absorptive contributions, both from the imaginary parts of the
self energies (see as the difference between the ``full'' and ``full R'' curves),
as well as from the imaginary parts of the vertex corrections,
turn out to be of a few percent. 
On the contrary, 
in  scenario $\SE$, where only 
  $\DecayCmNW{2}{1,2}$ is kinematically allowed,
the mixing of the neutralinos is small, 
and consequently the dependence on $\phiMe$ turns out to be much smaller.
The size of the one-loop corrections, 
 reach \order{10\%} for $\SZ$ and show an important dependence on $\phiMe$.
For $\SE$ the corrections are of the order of 
 a few percent with a negligible $\phiMe$ dependence.

%%%%%%%%%%%%%%%%%%%%%%%%% F I G U R E %%%%%%%%%%%%%%%%%%%%%%%%%%%%%%%%%%%%%%%%%
%%%%%%%%%%%%%%%%%%%%%%%%% F I G U R E %%%%%%%%%%%%%%%%%%%%%%%%%%%%%%%%%%%%%%%%%

\begin{figure}[tb!]
\begin{center}
\begin{tabular}{c}
\includegraphics[width=0.49\textwidth,height=4.6cm]{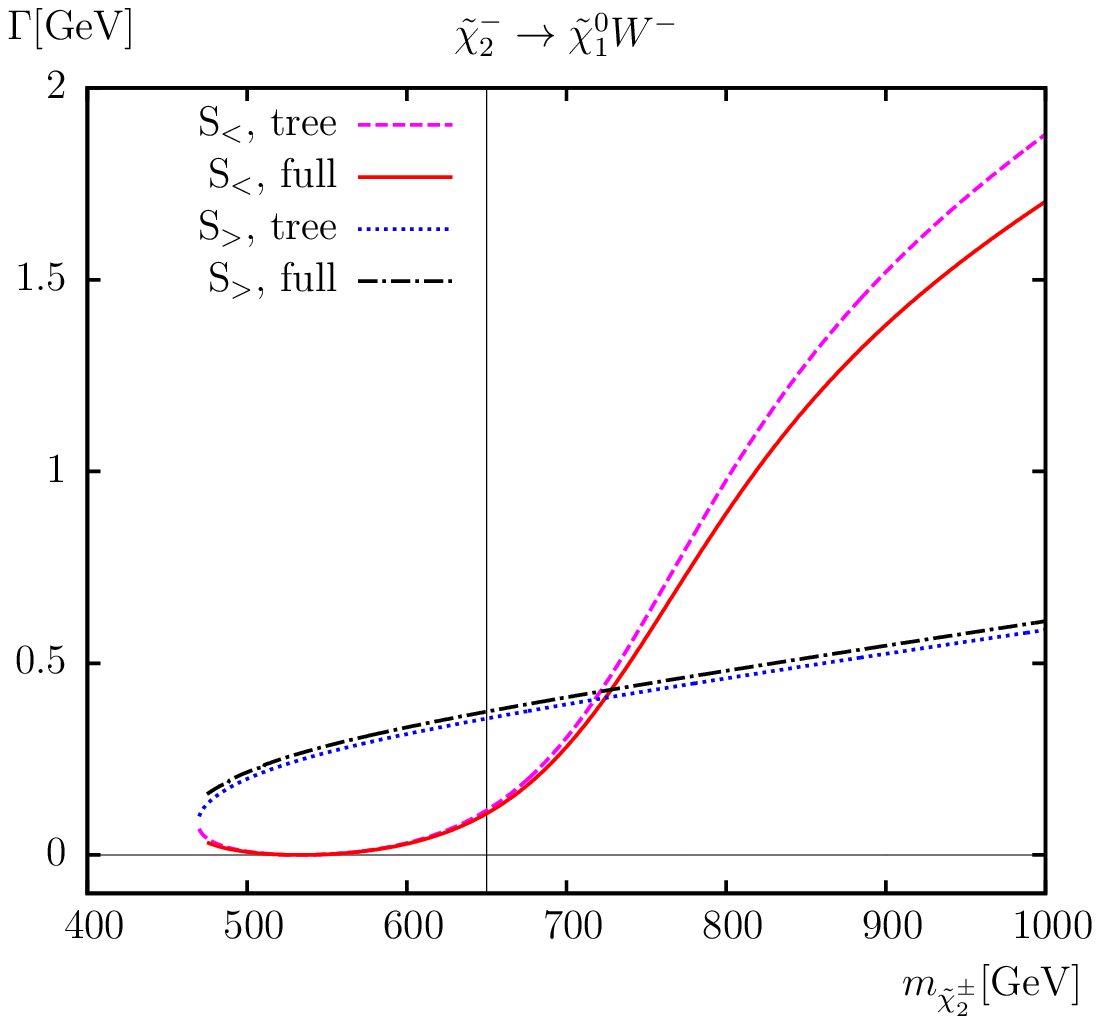}
\hspace{-4mm}
\includegraphics[width=0.49\textwidth,height=4.6cm]{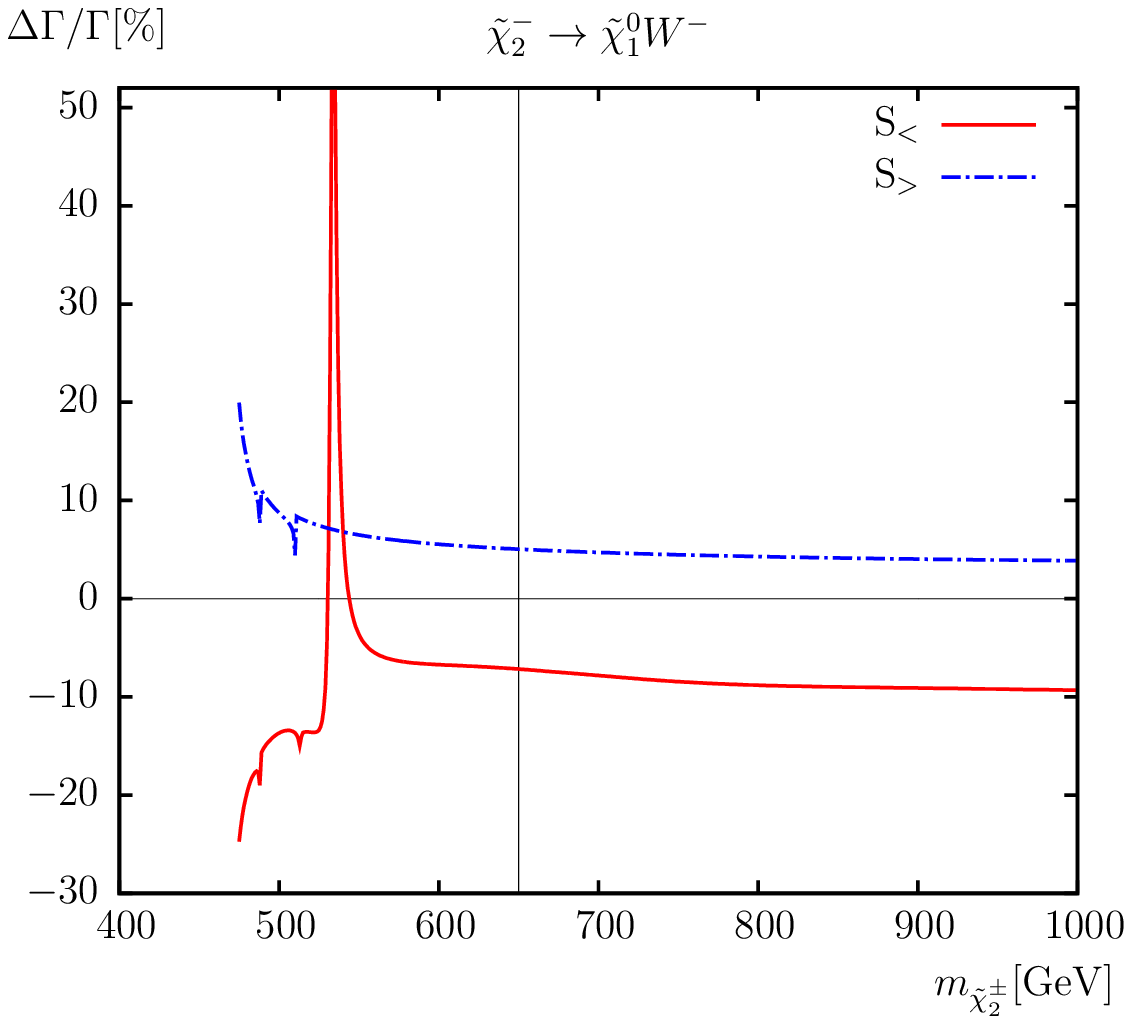} 
\end{tabular}
%\vspace{2em}
\begin{tabular}{c}
\includegraphics[width=0.49\textwidth,height=4.6cm]{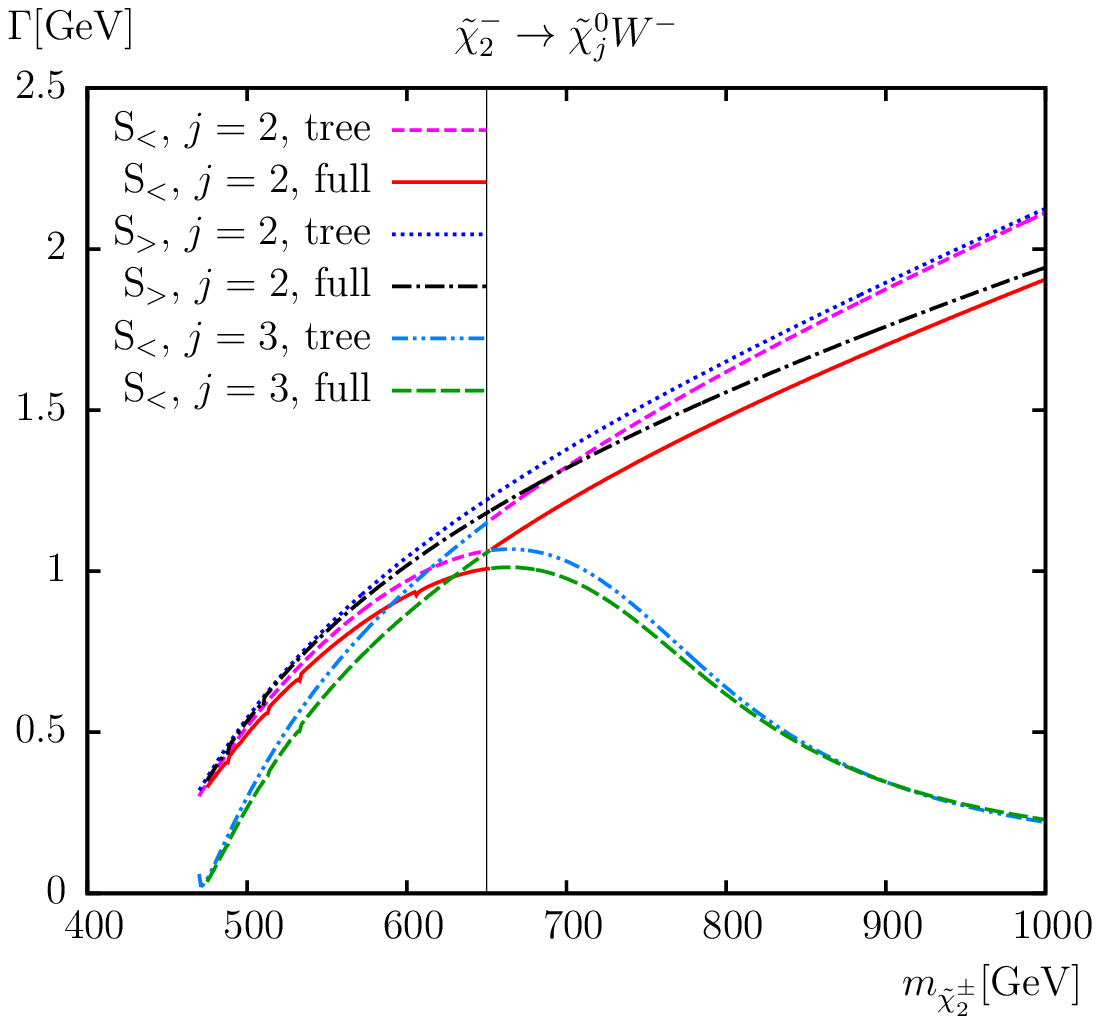}
\hspace{-4mm}
\includegraphics[width=0.49\textwidth,height=4.6cm]{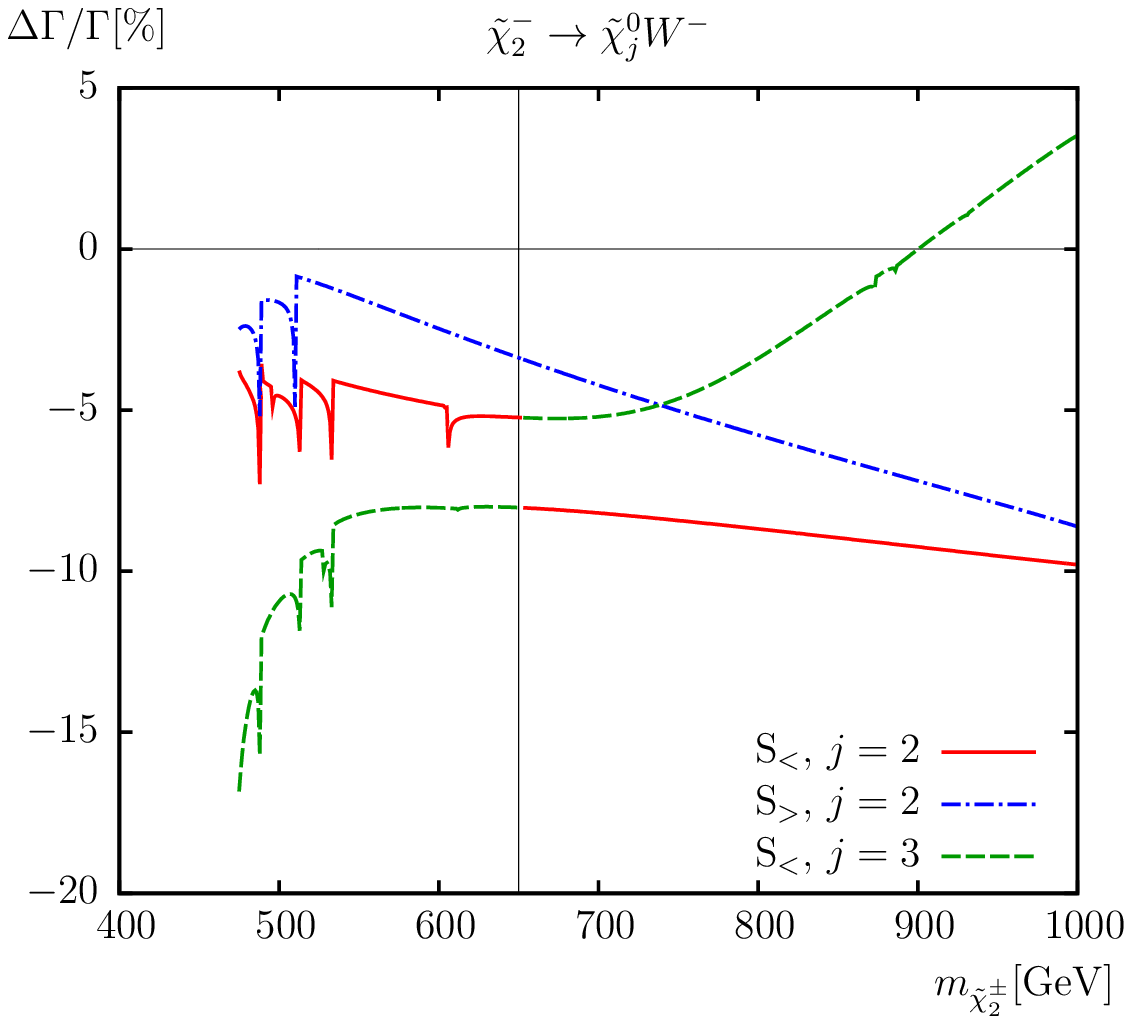} 
\end{tabular}
%\vspace{-2em}
\caption{
  $\Ga(\DecayCmNW{2}{j})$, $j=1,2,3$.
  Tree-level (``tree'') and full one-loop (``full'') corrected 
  decay widths are shown with the parameters chosen according to \SN\
  (see \refta{tab:para}), with $\mcha{2}$ varied.
  The left (right) plots show the decay width
  (the relative size of the corrections).
}
\label{fig:mC.cha2neujw}
\end{center}
\end{figure}

Figure \ref{fig:mC.cha2neujw} shows 
  $\Ga(\DecayCmNW{2}{1})$ (top) and
  $\Ga(\DecayCmNW{2}{2,3})$ (bottom row)
as a function of $\mcha{2}$, keeping all other parameters as in 
 \refta{tab:para}. As in the previous figure, the left (right) column shows 
 (the relative size of the corrections of) the decay widths.
  The vertical lines indicate where $\mcha{1} + \mcha{2} = 1000 \gev$, 
  i.e.\ the maximum reach of the ILC(1000).
Coincidentally, around this value of $\mcha{2}$ we observe a level crossing of the 
second and third neutralino in $\SZ$, which points out at the large neutralino mixing for these paramters.

The decay widths show a strong dependence on $\mcha{2}$ 
which is mainly due to the changing chargino-neutralino couplings, 
as well as the change in the phase space.
For $\SZ$ the decay width into the lightest neutralino almost vanishes at one point,
resulting in large relative corrections.
The relative one-loop corrections are mostly of \order{10\%}.
The dips in the one-loop corrections  are due to thresholds in
the vertex corrections.
It should be noted that a calculation very close to threshold requires
the inclusion of additional (non-relativistic) contributions, which is
far beyond the scope of this analysis. 

The decay width into $\neu{2}$ reaches $\sim 2\gev$ at $\mcha{2}=1\tev$,
while for the decay into $\neu{3}$ it reaches $\sim 1\gev$ in the region of maximal neutralino mixing.
The relative one-loop corrections are  of \order{5-10\%}.

%%%%%%%%%%%%%%%%%%%%%%%%% F I G U R E %%%%%%%%%%%%%%%%%%%%%%%%%%%%%%%%%%%%%%%%%
%%%%%%%%%%%%%%%%%%%%%%%%% F I G U R E %%%%%%%%%%%%%%%%%%%%%%%%%%%%%%%%%%%%%%%%%

Summarizing, we reviewed the evaluation of two-body decay modes of charginos in the cMSSM,
and show numerical results for the decay of the heavier chargino into neutralinos.
The relative size of the one-loop corrections is found  to be significant 
and should be taken into account in a reliable determination of the chargino/neutralino sector parameters.
This also applies in particular to the effects of the imaginary parts of the self-energies of the external particles.

% %%%%%%%%%%%%%%%%%%%%%%%%%%%%%%%%%%%%%%%%%%%%%%%%%%%%%%%%%%%%%%%%%%%%%%%%%%%%%

\section{Acknowledgments}

The work of S.H.\ was partially supported by CICYT (grant FPA
2007--66387 and FPA 2010--22163-C02-01).
F.v.d.P.\ was supported by 
the Spanish MICINN's Consolider-Ingenio 2010 Programme under grant MultiDark CSD2009-00064.

% ****************************************************************************
% BIBLIOGRAPHY AREA
% ****************************************************************************

\begin{footnotesize}
% IF YOU DO NOT USE BIBTEX, USE THE FOLLOWING SAMPLE SCHEME FOR THE REFERENCES
% ----------------------------------------------------------------------------

% ----------------------------------------------------------------------------

\end{footnotesize}

% ****************************************************************************
% END OF BIBLIOGRAPHY AREA
% ****************************************************************************

\end{document}